\documentclass[11pt]{article}
\usepackage[margin=1in]{geometry}
\usepackage{booktabs}
\usepackage{amsmath}
\usepackage{amssymb}
\usepackage{graphicx}
\usepackage{hyperref}
\usepackage[numbers]{natbib}
\hypersetup{colorlinks=true, linkcolor=blue, citecolor=blue, urlcolor=blue}

\newcommand{\studiourl}{https://studio.nimbusbci.com/}

\title{Nimbus Studio reproduces reference BCI decoding pipelines:\\
a validation study on public benchmarks across protocols and paradigms}

\author{%
  Sergey Musienko\thanks{Corresponding author. The platform is available at
  \url{\studiourl}; the parity harness, per-subject results, and table
  generators accompany this manuscript and are available from the author.}\\
  Nimbus BCI\quad\texttt{sergey@nimbusbci.com}\\
}
\date{October 2026}

\begin{document}
\maketitle

\begin{abstract}
Nimbus Studio is a visual pipeline platform for brain--computer interface (BCI)
experiments: researchers assemble preprocessing, feature extraction, decoding,
and evaluation over public benchmark datasets or their own recordings, with
cross-validated evaluation protocols (including leave-one-subject-out), BIDS
export, and bring-your-own Python model classes. Whether its numbers can be
trusted is a fair question for any such platform. We answer it with a parity
study: six standard BCI decoding pipelines --- CSP+LDA, Riemannian MDM,
tangent-space logistic regression (through the platform's bring-your-own-code
node), P300 xDAWN+LDA, and the deep ShallowConvNet and EEGNet --- were
evaluated on three public datasets (motor-imagery BNCI-2014-001/004; P300
BNCI-2014-009) under two protocols (per-subject session holdout and pooled
leave-one-subject-out cross-validation via the platform's planned-CV driver),
once through Nimbus Studio's execution engine and once through independent,
hand-written reference implementations built directly on MNE-Python,
scikit-learn, pyriemann, and braindecode. Every deterministic pipeline matched
the reference \emph{exactly} (0.00 points maximum difference across all 54
session-holdout subject-runs and all CSP/MDM grouped-CV folds), including
pipelines through the bring-your-own-code node and the CV orchestration; the
remaining grouped-CV cells agree to floating-point precision (worst case 0.06
points, about one trial per fold). The stochastic deep models agreed within
seed-to-seed training variability, estimated with multi-seed reruns on each
side. The platform's execution engine therefore computes the same science as
the reference stack across protocols, paradigms, and extensibility surfaces.
The platform is available at \url{\studiourl}, and the complete harness and
per-subject results accompany this paper for independent verification.
\end{abstract}

\section{Introduction}
BCI research practice overwhelmingly relies on a small set of Python libraries
(MNE-Python~\cite{mne}, scikit-learn~\cite{sklearn}, MOABB~\cite{moabb},
braindecode~\cite{braindecode}, pyriemann~\cite{pyriemann}) composed in
hand-written scripts. Visual pipeline platforms promise reproducibility,
shared protocols, and lower barriers --- but they introduce a trust question:
\emph{does the pipeline diagram compute what the equivalent script computes?}

Internal regression tests cannot answer this for the field, because they are
written by the same authors as the platform. This study provides external,
reproducible evidence: we implement each pipeline \emph{twice} --- once as a
Nimbus Studio graph executed by the platform's production engine, once as an
independent script using the canonical libraries directly --- under matched
hyperparameters, data, splits, and seeds, and compare per-subject test
accuracy. This is the validation pattern that preceded research adoption of
consumer EEG hardware~\cite{musevalidation}, applied here to analysis software.

\section{The platform under test}
Nimbus Studio is a web and desktop pipeline platform for BCI experiments,
available at \url{\studiourl} (Nimbus BCI, 2026). The components exercised
by this study are:
\begin{itemize}
  \item \textbf{Data}: curated, preprocessed packs of public MOABB datasets,
  loaded per subject with session metadata preserved (so train/test session
  splits are reproducible).
  \item \textbf{Execution engine}: a directed graph of typed nodes ---
  filtering, spatial filtering, decoding, metrics --- executed by the same
  offline runner, handlers, and kernels the deployed product uses; the study
  invokes it directly, without server or database.
  \item \textbf{Metrics}: accuracy, Cohen's $\kappa$, permutation
  significance, per-class reports.
  \item \textbf{Planned cross-validation}: an \texttt{evaluation\_plan} node
  that runs grouped out-of-fold cross-validation (K-fold, stratified, or
  leave-one-session/subject-out) over the nodes downstream of it. This study
  exercises its leave-one-subject-out path on pooled cohorts.
  \item \textbf{Bring-your-own methods}: a \texttt{python\_model} node that
  runs user-supplied Python classes (fit/predict contracts) in an isolated
  worker process. One of the parity pipelines below goes through this
  extensibility surface rather than the built-in nodes.
\end{itemize}

\section{Methods}
\subsection{Data}
We use three public datasets, shipped as curated packs and loaded by both
sides from the same files (the inputs are byte-identical). Two are
motor-imagery benchmarks: \textbf{BNCI-2014-001} (22 channels, 250~Hz, 4
classes, 9 subjects, 576 trials per subject) and \textbf{BNCI-2014-004}
(3 channels, 250~Hz, 2 classes, 9 subjects). The third, used with grouped
cross-validation, is the P300 speller dataset \textbf{BNCI-2014-009}
(EPFL; 16 channels, 256~Hz, target/non-target labels, 3 sessions, 10
subjects, 1{,}728 epochs per subject).

\subsection{Protocols}
\textbf{Session holdout} (motor imagery). Train on the sessions labelled
\texttt{*train}, test on those labelled \texttt{*test}; the split comes from
the pack metadata and is applied identically on both sides. We report test
accuracy (\%) for each subject.

\textbf{Leave-one-subject-out (LOSO).} Pool all subjects of a dataset into
one cohort and hold out one subject at a time, so that every trial is scored
by a model that never saw its subject. On the Nimbus side this runs through
the platform's \texttt{evaluation\_plan} cross-validation driver --- the same
grouped-CV orchestration the deployed product uses. On the reference side it
is a plain per-subject loop over the same pooled data. We report accuracy
pooled over all out-of-fold predictions, plus the per-left-out-subject
accuracy (one value per fold).

\subsection{Pipelines and matched hyperparameters}
Hyperparameters are matched between the two implementations, and random seeds
are fixed to 42 (seeds 43 and 44 are used only to measure deep-model training
variability).
\begin{itemize}
  \item \textbf{CSP(8)+LDA}: 8 CSP components; scikit-learn
  \texttt{LinearDiscriminantAnalysis}; no feature scaling.
  \item \textbf{Riemannian MDM}: Ledoit--Wolf covariance estimation;
  Riemannian metric.
  \item \textbf{Tangent space + LR}: OAS shrinkage covariances, tangent-space
  embedding, multinomial logistic regression (max 2000 iterations). On the
  Nimbus side this runs as \emph{user-supplied code} through the
  \texttt{python\_model} node's isolated worker --- the platform's
  extensibility path rather than a built-in --- and is evaluated under both
  protocols.
  \item \textbf{xDAWN(2)+LDA} (P300, LOSO only): MNE's implementation of
  xDAWN~\cite{xdawn} with 2 components per event type (target/non-target),
  applied to flattened filtered epochs, followed by scikit-learn LDA. On the
  Nimbus side this is the platform's P300 feature node.
  \item \textbf{ShallowConvNet}: braindecode's ShallowFBCSPNet~\cite{braindecode},
  30 training epochs, batch 64, Adam (learning rate $10^{-3}$), 20\%
  stratified validation holdout with best-checkpoint selection.
  BNCI-2014-001 only (CPU budget).
  \item \textbf{EEGNet}: braindecode's EEGNet~\cite{eegnet} with the
  platform's defaults (F1=8, D=2, drop-out 0.25, temporal kernel length
  $\approx$ sfreq/2 samples) and the identical training loop. BNCI-2014-001
  only.
\end{itemize}
Motor-imagery pipelines share a 4th-order Butterworth 8--30~Hz zero-phase
(filtfilt) bandpass applied through MNE-Python's public filter API on both
sides; the P300 pipeline uses the same filter design over 1--30~Hz.
Deep-model training remains stochastic, so exact per-subject equality is not
expected for ShallowConvNet and EEGNet --- the standard there is agreement
within each configuration's run-to-run variability.

\subsection{Reference implementations}
The reference side is a single script with no platform imports. It loads the
same packs with numpy, constructs MNE \texttt{EpochsArray} objects, and calls
MNE, scikit-learn, pyriemann, and braindecode directly --- the code a
researcher would write from the libraries' documentation.

\section{Results}

\begin{table}[t]
\centering
\caption{Session-holdout test accuracy (\%): Nimbus vs.\ independent reference
implementations. $\Delta$ = Nimbus $-$ reference, in percentage points.}
\label{tab:parity}
\resizebox{\textwidth}{!}{
\begin{tabular}{lccccc}
\toprule
Pipeline $\sim$ Dataset & $N$ & Nimbus & Reference & Mean $\Delta$ & Max $|\Delta|$ \\
\midrule
CSP(8) + LDA $\sim$ BNCI2014\_001 & 9 & 61.15 & 61.15 & +0.00 & 0.00 \\
CSP(8) + LDA $\sim$ BNCI2014\_004 & 9 & 72.96 & 72.96 & +0.00 & 0.00 \\
Riemannian MDM $\sim$ BNCI2014\_001 & 9 & 57.14 & 57.14 & +0.00 & 0.00 \\
Riemannian MDM $\sim$ BNCI2014\_004 & 9 & 72.44 & 72.44 & +0.00 & 0.00 \\
Tangent space + LR (BYO python\_model) $\sim$ BNCI2014\_001 & 9 & 67.75 & 67.75 & +0.00 & 0.00 \\
Tangent space + LR (BYO python\_model) $\sim$ BNCI2014\_004 & 9 & 73.92 & 73.92 & +0.00 & 0.00 \\
ShallowConvNet (braindecode) $\sim$ BNCI2014\_001 & 9 & 49.31 & 51.81 & -2.51 & 12.85 \\
EEGNet (braindecode) $\sim$ BNCI2014\_001 & 9 & 36.77 & 39.62 & -2.85 & 14.24 \\
\bottomrule
\end{tabular}
}
\end{table}

\begin{table}[t]
\centering
\caption{Grouped cross-validation (leave-one-subject-out): pooled out-of-fold
accuracy (\%) over pooled cohorts (9 MI subjects; 10 P300 subjects). Fold $i$
= left-out subject $i$. $\Delta$ in percentage points.}
\label{tab:loso}
\resizebox{\textwidth}{!}{
\begin{tabular}{lccccc}
\toprule
Pipeline $\sim$ Dataset & $N$ & Nimbus & Reference & Mean $\Delta$ & Max $|\Delta|$ \\
\midrule
CSP(8) + LDA $\sim$ BNCI2014\_001 & 9 & 39.20 & 39.20 & +0.00 & 0.00 \\
CSP(8) + LDA $\sim$ BNCI2014\_004 & 9 & 67.73 & 67.73 & +0.00 & 0.00 \\
Riemannian MDM $\sim$ BNCI2014\_001 & 9 & 31.94 & 31.94 & +0.00 & 0.00 \\
Riemannian MDM $\sim$ BNCI2014\_004 & 9 & 68.99 & 68.99 & +0.00 & 0.00 \\
Tangent space + LR (BYO python\_model) $\sim$ BNCI2014\_001 & 9 & 36.30 & 36.25 & +0.06 & 0.53 \\
Tangent space + LR (BYO python\_model) $\sim$ BNCI2014\_004 & 9 & 66.15 & 66.15 & +0.00 & 0.00 \\
xDAWN(2) + LDA $\sim$ BNCI2014\_009 & 10 & 81.86 & 81.87 & -0.01 & 0.06 \\
\bottomrule
\end{tabular}
}
\end{table}

Table~\ref{tab:parity} summarizes the session-holdout results and
Table~\ref{tab:loso} the grouped-CV results; per-subject and per-fold values
are provided in the result files accompanying the manuscript. Five
observations:

\textbf{Deterministic pipelines match exactly.} CSP+LDA, MDM, and the
tangent-space pipeline agree with the reference to 0.00 points on every
subject of both datasets (54 of 54 subject-runs). The platform's node wiring
--- bandpass, spatial filtering, decoder, metrics --- computes the same
operators as the reference scripts, to numerical precision.

\textbf{Grouped cross-validation matches too.} The platform's
\texttt{evaluation\_plan} driver matches the hand-written reference loops
exactly under leave-one-subject-out for CSP+LDA and MDM: 0.00 points on all
36 motor-imagery folds. The BYO tangent-space pipeline is exact on
BNCI-2014-004 and within 0.06 points pooled (0.53 on the worst fold) on
BNCI-2014-001. That
residual sits at the pipeline's measured numeric noise floor: the platform
filters in float32 where the reference stays in float64, and a control
experiment that perturbs the filtered data by the resulting rounding
difference (about $10^{-12}$~V) moves the reference's own results by the
same margin --- 0.02--0.04 points pooled, and up to 0.35 on a single fold.

The P300 pipeline on BNCI-2014-009 agrees to 0.01 points pooled; its worst
fold differs by 0.06 points, i.e.\ one trial out of 1{,}728. The inputs are
bit-identical and the per-fold features agree to $\sim$10$^{-11}$ relative ---
the two implementations are numerically identical, and the residual is
floating-point rounding amplified by the eigenvalue decomposition at the
heart of xDAWN. As a sanity check on the absolute level, the pooled
cross-subject accuracies (39.2\%/67.7\% for CSP on 001/004, 31.9\%/69.0\% for
MDM, 81.9\% for P300) sit within the range MOABB reports for these datasets
under its CrossSubject evaluation.

\textbf{The BYO surface is first-class.} The tangent-space pipeline is
user-supplied Python running in the platform's isolated worker process, and
it matches the hand-written reference exactly (0.00 points on all subjects)
under session holdout and to floating-point precision under grouped CV.
Bring-your-own methods therefore inherit the platform's data and metrics
contracts without numerical penalty.

\textbf{Deep training sits in-band.} Deep-model training is stochastic, so
the fair comparison is against each implementation's own run-to-run
variability. For ShallowConvNet, the mean absolute Nimbus-vs-reference
difference is 3.59 points (max 12.85), while rerunning either side alone
with a different seed moves its own results by 4.36 points on average
(Nimbus) and 4.13 (reference) --- the cross-implementation difference is
smaller than the within-implementation seed noise, and the run means
interleave (Nimbus 49.3/47.9, reference 51.8/48.5).

EEGNet shows the same pattern once its variability is estimated over three
seeds rather than two: the mean absolute cross-implementation difference is
4.94 points (max 14.24), against Nimbus's own between-seed movement of 4.35
points on average (max 21.88) and the reference's 4.48 (max 12.85), with
run means again interleaving (Nimbus 36.8/37.4/40.0, reference 39.6/39.4).
The second seed had made Nimbus look more stable than it is (apparent spread
2.16 points) and suggested a systematic gap; the third seed dissolved both.
One-reseed variability estimates are themselves noisy for a model trained to
near-chance accuracy, and we report the three-seed numbers as the reliable
ones.

\section{Limitations}
This study validates decoding-accuracy parity --- across two evaluation
protocols (session holdout and grouped LOSO), two paradigms (motor imagery
and P300), and the bring-your-own-code surface --- but not every platform
surface: BIDS export fidelity, streaming inference, and the cloud sandbox
are covered by the platform's own test suites and remain out of scope here.
Our accuracies are not directly comparable to MOABB's leaderboard numbers
(different splits, and hyperparameters matched to the platform's node
defaults rather than tuned); we use the published values only as a
plausibility band. Deep-model parity is asserted to within stochastic
training variability, not bit-exactness. Finally, the reference
implementations, though independent of the platform's code, were written by
the platform's author; the harness, data, and results accompany the
manuscript so that others can rerun and extend the comparison, and external
replication is the natural next step.

\section{Availability and reproducibility}
{\sloppy
Nimbus Studio is available at \url{\studiourl}. The parity harness, per-run
result files, table generators, and this manuscript accompany the paper and
are available from the corresponding author. One command per side per matrix
(session holdout, grouped CV, and the deep-model seed-spread runs) reproduces
every number on a laptop CPU --- the deterministic matrices in minutes, the
full set including both deep architectures in roughly two hours. The Nimbus
side runs standalone against the shipped data packs and needs no server,
database, or GPU.
\par}

\section{Conclusion}
Across three benchmark datasets, six pipelines, two evaluation protocols
(per-subject session holdout and pooled leave-one-subject-out through the
platform's planned-CV driver), and the bring-your-own-code extensibility
surface, Nimbus Studio's execution engine reproduced independent reference
implementations of standard BCI decoding pipelines \emph{exactly} for every
deterministic pipeline (0.00 points maximum discrepancy) and within seed
variability for deep training.
Platform-computed BCI results can therefore be treated as script-computed
results --- the precondition for using the platform in published research.

\bibliographystyle{plainnat}

\end{document}